\documentstyle[epsfig]{aipproc}

\begin{document}

\hspace*{\fill}\parbox[b]{3.4cm}{MSUHEP-70624 \\ June 1997\\
hep-ph/9706496\\
\vspace*{1.5cm}}

\title{QCD Phenomenology of Charm Production at HERA
\thanks{Presented at the $5^{\rm th}$ International Workshop on Deep Inelastic
Scattering and QCD (DIS 97), Chicago, IL, USA, April 14-18, 1997.  Work done 
in collaboration with Jim Amundson, Wu-Ki Tung, and Xiaoning Wang.}}

\author{Carl R. Schmidt}
\address{Department of Physics and Astronomy\\
Michigan State University\\
East Lansing, MI 48824, USA}

\maketitle

\begin{abstract}

We compare different schemes for the treatment of heavy 
quark production in Deep-Inelastic Scattering (DIS).  For fully-integrated 
quantities such as $F_{2}(x,Q^{2})$, we advocate the use of the 
General-Massive Variable-Flavor-Number (GM-VFN) scheme; we present 
some results showing the progress of a Next-to-Leading Order calculation 
in this scheme.  For differential quantities, the Fixed-Flavor-Number (FFN) 
scheme provides a more appropriate starting point.  We present a new 
calculation of 
the azimuthal distribution of charm quark production in DIS.  All 
results have been obtained using a Monte Carlo program under 
development. 

\end{abstract}

\subsection*{Introduction}

The theoretical treatment of heavy quark production has undergone much
re-examination recently.  A basic question is whether, or not, the charm quark
(or bottom quark) should be included in the initial state, with its own
parton density function (PDF).  Traditionally, there have been two approaches
used, which we shall call the Fixed-Flavor-Number (FFN) scheme and the 
Zero-Mass 
Variable-Flavor-Number (ZM-VFN) scheme.  In the FFN scheme one does not 
include charm as a 
parton in the nucleon.  The charm quark only occurs in the final state, and
its mass is treated exactly.  This is the standard approach for calculations
of specific heavy flavor production process, such as charm production in
deep inelastic scattering \cite{smith} and charm or bottom production in 
hadron-hadron colliders \cite{dawson}.  Alternatively, in the ZM-VFN scheme 
one does include a heavy quark PDF in the nucleon  at scales above 
the quark mass, but the quark is treated as massless in both the initial and 
final states.  This is the approach used in most other inclusive jet
calculations and parton shower Monte Carlo simulations.  It also corresponds
to the treatment of the charm and bottom quark in most global analyses of
the parton density functions, including DIS structure functions.

Given these two seemingly contradictory methods of calculating heavy quark
production cross sections, one may wonder which is correct.  The answer, of
course, is that both are correct at some level.  The real question is what is
the best method for calculating the particular set of observables
under consideration at the energy scale of the processes being studied.  For
example, near the charm production threshold in deep inelastic scattering,
the various energy scales, $Q^2\sim m_c^2\sim W^2$, are roughly equal, so 
that we have essentially a one scale problem.  Since the FFN scheme treats
this scale exactly, including the kinematics of the charm quark mass, it
should be expected to do well here.  However, at higher $Q^2$ scales 
the effect of the $m_c$ can be neglected, except when it occurs in large
logarithms $\log(Q^2/m_c^2)$.  These can be resummed by invoking a charm
quark parton density function and evolving up to $Q^2$.  Thus, at higher
energies the ZM-VFN scheme will be more accurate.  As a side remark, we also
note that the ZM-VFN scheme is necessary if one is to consider the possibility
of a nonperturbative charm quark contribution \cite{brodsky} to the nucleon
PDF beyond that which is generated dynamically by the splitting
of gluons.

In general we want a scheme that does the best job at
including the largest corrections from higher orders in perturbation theory.
Since much of the data is in an intermediate range of $Q^2$, neither
the FFN nor the ZM-VFN schemes are completely adequate.  For this purpose
a General-Massive Variable-Flavor-Number (GM-VFN) scheme was devised 
\cite{ACOT}.  (It is sometimes referred to as ACOT, for the originators.)
In this scheme the charm quark
PDF evolves with massless evolution, but the heavy quark mass is kept
everywhere else in the matrix elements and phase space.  The leading order
process $\gamma^*c\rightarrow c$ from the ZM-VFN scheme and the leading
order process $\gamma^*g\rightarrow c\bar c$ from the FFN scheme are
both included in the new scheme.   However, to keep from double-counting,
we must also subtract a term which is the convolution of the $g\rightarrow q$
splitting function with the $\gamma^*q\rightarrow q$ process.  This is 
shown diagrammatically in figure 1.  Note that, although the subtraction
term may appear ad hoc here, it is in fact just the massive equivalent to the
Altarelli-Parisi subtraction of the $1/\epsilon$ pole at next-to-leading
order (NLO) in the ZM-VFN scheme.

\begin{figure}[t!] 
\centerline{\epsfig{file=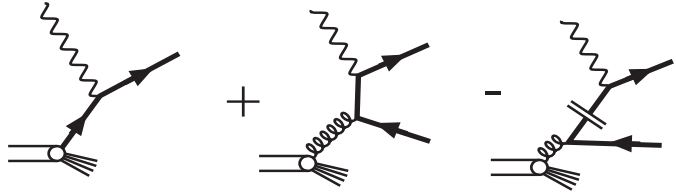,height=1.0in,width=3.5in}}
\vspace{10pt}
\caption{Feynman diagrams for LO GM-VFN scheme.}
\label{fig1}
\end{figure}

By including the three terms shown in figure 1 with massless evolution
of the charm PDF, but full mass-dependence in 
the coefficient functions, we find that the calculation in the GM-VFN scheme
agrees with the ZM-VFN scheme at large $Q^2$ and with the FFN scheme near
threshold.  The three terms in figure 1 constitute a Leading Order 
(LO) calculation if one
considers the charm quark PDF to be $O(\alpha_s)$.  This makes sense
in the absence of nonperturbative charm contributions since the charm 
quark PDF arises
entirely from splitting of gluons and is consequently smaller than the
gluon PDF.  Furthermore, this method of $\alpha_{s}$ counting leads to 
reduced  renormalization/factorization scale ($\mu$) dependence, as we 
shall see.  This scheme 
also  allows the inclusion of nonperturbative charm contributions to the 
nucleon PDF, which should exist at some level.  A
recently-proposed prescription by Martin {\it et al.} (MRRS) \cite{MRRS}
 is similar in spirit to this.  They
use the conventional counting of $\alpha_s$ and split the heavy quark 
mass-dependence into a contribution to the splitting functions and 
a contribution to the coefficient functions.   See the talk by Olness at 
this conference for a discussion of this issue \cite{olness}.

For a more detailed discussion of the theory of heavy quark 
production, we refer to the talk by Tung at this conference \cite{tung}.  
In the remainder of the present talk we describe the progress in implementing 
the GM-VFN scheme at next-to-leading order (NLO) in a Monte Carlo 
program.  Currently, we have added the ${\cal O}(\alpha_{s})$ virtual and 
real ($\gamma^*c\rightarrow gc$) contributions to the 
charm excitation process, with the appropriate subtraction.  This 
partial-NLO GM-VFN calculation
incorporates all of the terms that are in the NLO calculation in 
the ZM-VFN scheme using the conventional $\alpha_{s}$-counting.  The 
${\cal O}(\alpha_{s})$ virtual and real corrections 
to the gluon- or light quark-initiated processes (and subtractions) that also
contribute to the NLO GM-VFN scheme have not yet been included.  
However, we can still gain useful insight with the results so far. 

\subsection*{Integrated observables -- $F_2(x,Q^2)$}

As discussed above, the GM-VFN scheme should be the best method of 
calculating integrated observables, such as the charm contribution to 
$F_{2}(x,Q^{2})$.  In figures 2a and 2b we compare the calculation of 
$F_2(x,Q^2)$ in the various schemes. In figure 2a $F_2(x,Q^{2})$ 
is plotted as a function of $Q^2$ for $x=0.01$.  We see that
the NLO FFN calculation gives a sizeable increase over the LO FFN (tree-level
gluon fusion) calculation, while the LO GM-VFN scheme is even higher.  
The partial-NLO GM-VFN, which should be most accurate at high $Q^{2}$, 
gives a small correction over LO GM-VFN.  This indicates that the FFN tends 
to underestimate the cross section at high $Q^{2}$.  At intermediate 
$Q^{2}$ we will need the full-NLO GM-VFN calculation, which is underway.
Note that the GM-VFN scheme is much more economical than the the FFN
scheme, since the largest contributions to the cross section are included
at LO with substantially less work.

In figure 2b we show the $\mu$-dependence of $F_2(x,Q^{2})$ for $x=0.01$ and 
$Q=10$ GeV.
In this plot the NLO FFN calculation only gives a slight improvement in the
scale dependence over the LO FFN.  However, the LO GM-VFN calculation is
less sensitive to the choice of scale.
The partial-NLO GM-VFN calculation actually 
has larger $\mu$-dependence than LO GM-VFN, but the remaining terms 
in the NLO GM-VFN calculation should improve this again.  In both 
figures 2a and 2b the wiggles are just due to Monte Carlo statistics.

\begin{figure}[t!] 
\centerline{\epsfig{file=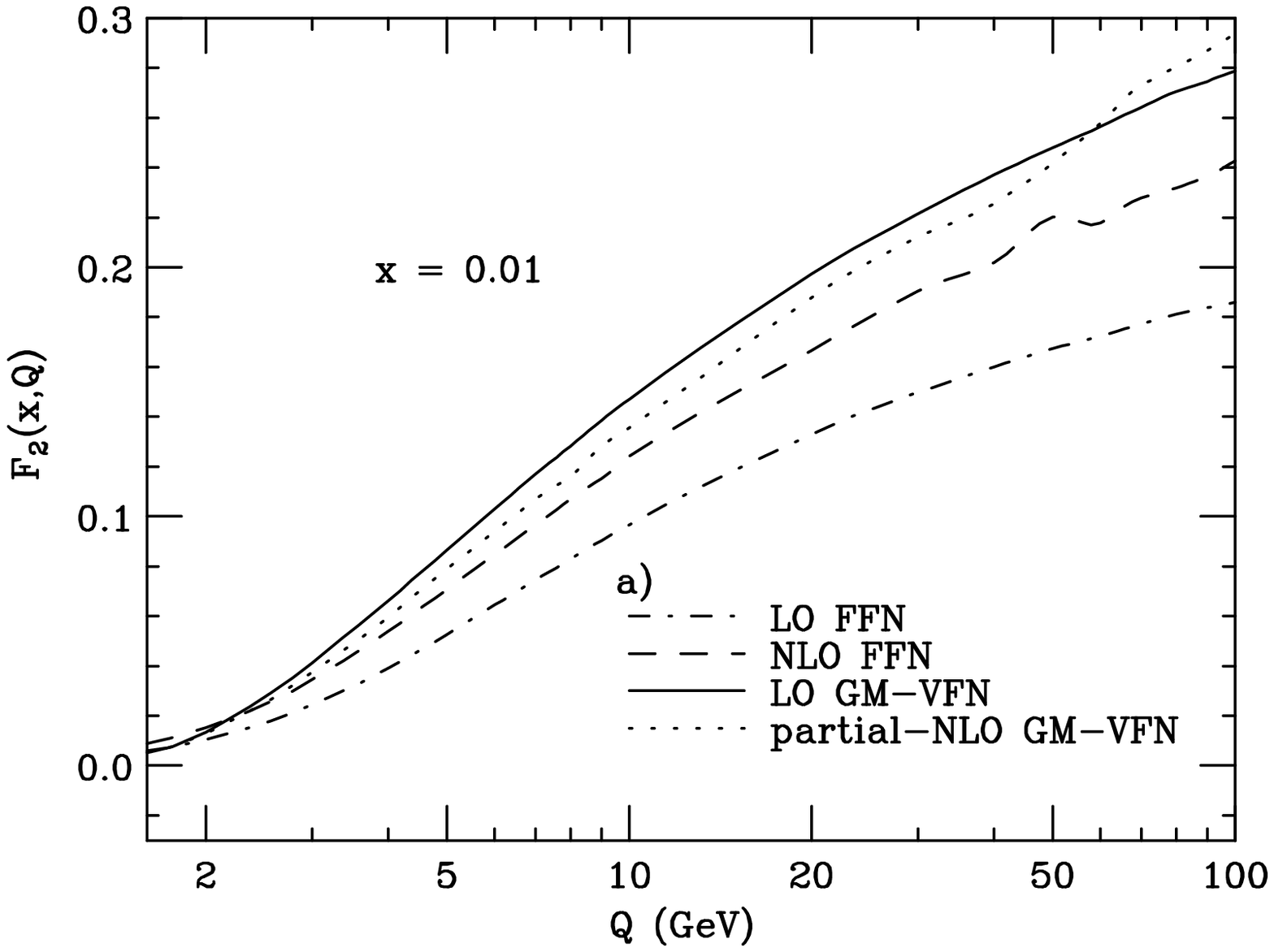,height=2.3in,width=3.0in}
\epsfig{file=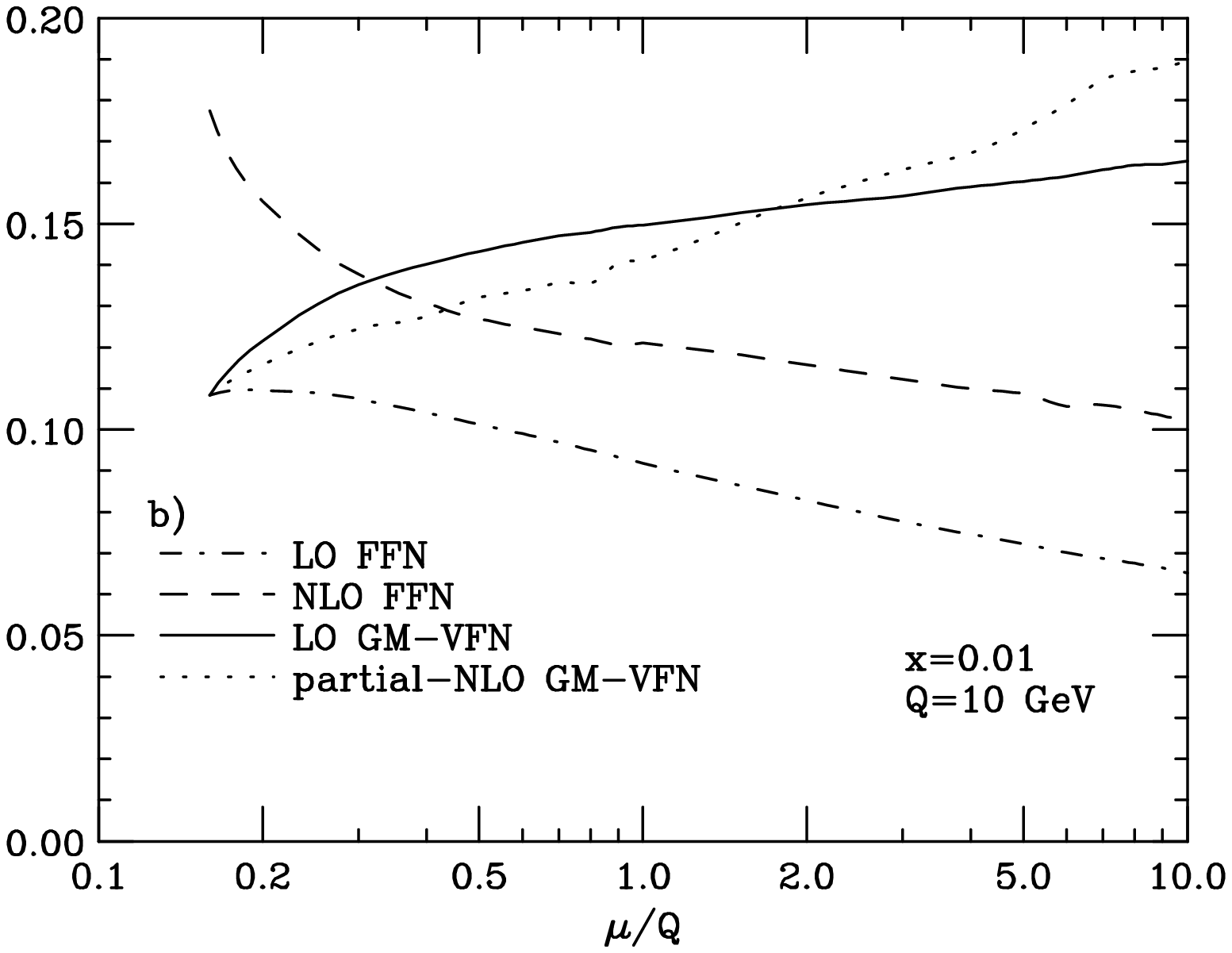,height=2.3in,width=2.8in}}
\vspace{10pt}
\caption{$F_{2}(x,Q^{2})$ in various schemes.
a) versus $Q$ (GeV). b) versus $\mu/Q$.  The
NLO FFN curves are calculated using the code of Harris and Smith [6].}
\label{fig2}
\end{figure}

\subsection*{Differential observables -- $\phi$ distribution}

There are issues that make the GM-VFN scheme somewhat 
problematic when applied to differential distributions.  The 
initial-state charm quark contributions include a resummation of
collinear radiation diagrams, with the phase space of the anticharm 
quark always integrated out.  Thus, the kinematics cannot be treated exactly.
For instance, the transverse momentum distribution with respect to the 
$\gamma^{*}P$-axis is a delta-function, which can only make sense upon 
integration.  In the FFN scheme, however, the $c\bar c+X$ final state 
is always treated exactly to a given order.  Therefore, we expect that 
it should give a reasonable prediction for the differential 
distributions, as long as we are not too far from threshold.

In figure 4 we present a new plot, using our Monte Carlo in the FFN 
scheme at LO, of the distribution of the azimuthal angle $\phi$.  This 
angle is defined as the angle between the lepton ($e\rightarrow e^{\prime}$) 
plane and the hadron ($c\bar c$) plane in a frame in which the 
$\gamma^{*}$ and the proton $P$ are collinear.  By symmetry the 
distribution will be of the form $d\sigma/d\phi\sim 
A+B\cos{\phi}+C\cos{2\phi}$.  In the plot we have made the cuts 
$0.01<y<0.7$, $10<Q^{2}<100$ GeV, and $p_{\perp}^{*}>2$ GeV, where 
$p_{\perp}^{*}$ is the transverse momentum with respect to the 
$\gamma^{*}P$-axis.  We have also summed over the charm and anticharm 
contributions, which removes the $\cos{\phi}$ term.  Although lab 
frame cuts will alter this distribution, they can easily be included 
in the Monte Carlo program.  

\begin{figure}[t!] 
\centerline{\epsfig{file=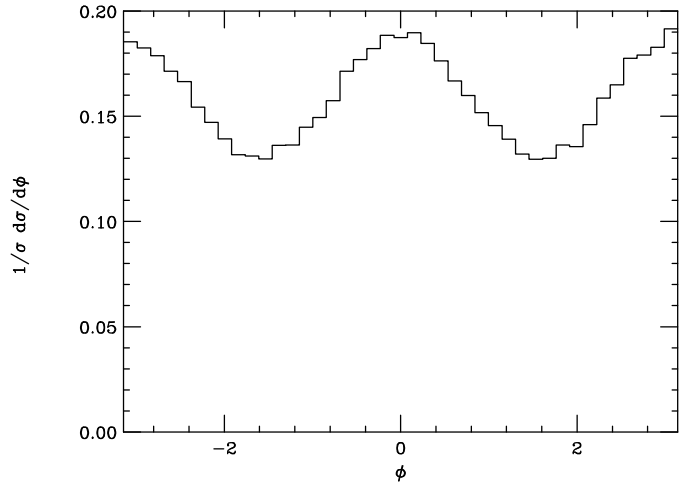,height=2.5in,width=3.5in}}
\vspace{10pt}
\caption{Distribution in the azimuthal angle $\phi$.}
\label{fig4}
\end{figure}


We would like to thank Bryan Harris \cite{harris} for providing us with 
the code to produce the NLO FFN plots.

\end{document}